\documentclass[12pt]{article}

\usepackage{amssymb,amsmath,amsthm}
\usepackage{array,longtable}
\usepackage[dvips]{graphicx}
\usepackage[dvips]{color}
\usepackage[mathscr]{eucal}

\newcommand{\finkfile}{}

\newcolumntype{V}{>{$}m{4cm}<{$}}
\newcolumntype{C}{>{$}c<{$}}
\newcolumntype{L}{>{$}l<{$}}
\newcolumntype{R}{>{$}r<{$}}

\newcommand{\Nc}{\mathcal{N}}

\newcommand{\arctanh}{\mathop{\mathrm{arctanh}}\nolimits}

\allowdisplaybreaks[3]

\renewcommand{\theequation}{\arabic{section}.\arabic{equation}}

\begin{document}
\title{
$~$
\vspace{-25mm}
\begin{flushright}
{
\small
\hfill hep-th/0210199
\\}
\end{flushright}
\vspace{5mm}
\textbf{Representation of Small Conformal}
\\
\textbf{Algebra in $\kappa$-basis}
$~$\\
$~$\\}
\author{
\textsf{D.M.~Belov}\footnote{On leave from Steklov Mathematical
Institute, Moscow, Russia.}
\vspace{3mm}
\\
Department of Physics
\\
Rutgers University
\\
136 Frelinghuysen Rd.
\\
Piscataway, NJ 08854, USA
\vspace{3mm}
\\
\texttt{belov@physics.rutgers.edu}
}

\date{~}
\maketitle
\thispagestyle{empty}

\begin{abstract}
In hep-th/0202087 it was argued that the operator $L_0$ is
bad defined in $\kappa$-basis as a kernel operator.
Indeed, we show that $L_0$ is a difference operator.
We also find a representation of $L_1$ and $L_{-1}$
in a class of difference operators.
\end{abstract}

\newpage


\section{Introduction}\label{sec:intro}
\dopage{\finkfile}
\setcounter{equation}{0}
The basic ingredients in the construction of the covariant
string field theory are Witten's star product \cite{Witten}
and BRST operator. Recent progress in the diagonalization
\cite{spectroscopy}
of the Neumann matrices $M^{\prime\,rs}$ defining star product
allows one to identify this product with the continuous Moyal
product \cite{gm,BK1}.
We have
$$
\sum_{n=1}^{\infty}M^{\prime\, rs}_{mn}v_{n}^{(\kappa)} =
\mu^{rs}(\kappa)v_{m}^{(\kappa)}
$$
where $-\infty<\kappa<\infty$, the eigenvalues $\mu^{rs}(\kappa)$ are
\begin{equation*}
\mu^{rs}(\kappa)=\frac{1}{1+2\cosh\frac{\pi\kappa}{2}}
\Bigl[1-2\delta_{r,s}+e^{\frac{\pi\kappa}{2}}\delta_{r+1,s}
+e^{-\frac{\pi\kappa}{2}}\delta_{r,s+1}\Bigr].
\end{equation*}
The eigenvectors $v_{n}^{(\kappa)}$ are given by their generating function
\begin{equation} \label{gen_fn}
f^{(\kappa)}(z)=\sum_{n=1}^{\infty}\frac{v_{n}^{(\kappa)}}{\sqrt{n}}z^{n} =
\frac{1}{\kappa\sqrt{{\cal N}(\kappa)}}(1-e^{-\kappa\tan^{-1}z})
\end{equation}
where
$$
{\cal N}(\kappa)=\frac{2}{\kappa}\sinh\left(\frac{\pi\kappa}{2}\right) \, .
$$

In spite of the fact that we understand the
nature of the star product, it is still unclear how BRST operator
acts in the basis in which star product is simple.
Moreover we do not know a representation of Virasoro generators in this basis.
The authors of \cite{gm}
tried to construct generator $L_0$ us a kernel operator in the $\kappa$-basis. They
show that this kernel is \textit{not} defined as a distribution.
In this paper I show that small conformal
algebra $\{L_{-1},\,L_0,\,L_1\}$ is represented by a certain difference
operators in the $\kappa$-basis.

Let me briefly remind a construction of the Fock space at hand \cite{berezin}. In the
discrete basis we have creation and annihilation operators $a_n^{\dag}$
and $a_n$, $n=1,2,\dots$. Introduce the corresponding operators
in the continuous basis via
\begin{equation}
a_{\kappa}^{\dag}=\sum_{n=1}^{\infty}v_{n}^{(\kappa)}a_n^{\dag}
\quad\text{and}\quad
a_{\kappa}=\sum_{n=1}^{\infty}v_{n}^{(\kappa)}a_n.
\end{equation}
Let $f_1(\kappa),\dots, f_m(\kappa)$ be functions from Schwartz space,
then the states
\begin{equation}
a^{\dag}(f_1)\dots a^{\dag}(f_m)|0\rangle,
\quad\text{where}\quad
a^{\dag}(f)=\int_{-\infty}^{\infty}d\kappa\, f(\kappa)a^{\dag}_{\kappa}
\end{equation}
form a dense subset in this Fock space.
Since the operator
\begin{equation}
L_0=\sum_{n=1}^{\infty}na_n^{\dag}a_n
\end{equation}
contains only one creation and one annihilation operator it is completely
determined by specifying its action on one particle states. Therefore
we can translate the action of operator $L_0$ on states to its action
on functions from the Schwartz space via
\begin{equation}
[L_0,\,a^{\dag}(f)]=a^{\dag}(L_0[f]).
\label{LaLf}
\end{equation}

In the next section I show that
\begin{equation}
L_0[f](\kappa)=\frac{1}{4}\left[
\sqrt{\kappa(\kappa+i2^-)}\,f(\kappa+i2^-)
+\sqrt{\kappa(\kappa-i2^-)}\,f(\kappa-i2^-)
\right],
\label{mainL0}
\end{equation}
where $2^{-}=2-0$.
The operator \eqref{mainL0} is defined
on a class of holomorphic functions in the strip $ -2<\Im \kappa<2$.
These functions may have poles on the boundary, and
the term $i2^-$ shows how one approaches these poles. I.e. one
should approach
the pole on the line $\kappa+2i$ from the bottom and the
pole on the line $\kappa-2i$ from the top.
If it happens that the function $f$ is holomorphic in strip
$-4<\Im \kappa<4$ one can apply operator \eqref{mainL0} once again, etc.

Notice that the operators $L_1$ and $L_{-1}$ also contain only one creation
and one annihilation operator, and therefore they are completely
determined by specifying their action on one particle subspace. We can write
\begin{equation}
[L_{\pm 1},\,a^{\dag}(f)]=a^{\dag}(L_{\pm 1}[f]),
\end{equation}
where
\begin{subequations}
\begin{align}
L_1[f]&=-\frac{\kappa}{2}\,f(\kappa)+\frac{i}{4}
\bigl[\sqrt{\kappa(\kappa+i2^-)}\,f(\kappa+i2^-)
-\sqrt{\kappa(\kappa-i2^-)}\,f(\kappa-i2^-)\bigr]
\\
L_{-1}[g]&=-\frac{\kappa}{2}\,f(\kappa)-\frac{i}{4}
\bigl[\sqrt{\kappa(\kappa+i2^-)}\,f(\kappa+i2^-)
-\sqrt{\kappa(\kappa-i2^-)}\,f(\kappa-i2^-)\bigr]
\end{align}
\label{mainL1}
\end{subequations}

\bigskip
The paper is organized as follows. In Sections~\ref{sec:L0}
and \ref{sec:L1} we
give a derivation of the equations \eqref{mainL0} and \eqref{mainL1}
correspondingly. In Section~\ref{sec:com} we check that
the generators \eqref{mainL0} and \eqref{mainL1} defining in a
class of difference operators have indeed correct commutation
relations.
The Appendix contains the technical information.

\section{Operator $L_0$}
\label{sec:L0}
\setcounter{equation}{0}

\dopage{\finkfile}

Using the fact that
\begin{equation}
[L_0,a_n^{\dag}]=na_n^{\dag}.
\end{equation}
one can easily obtain that $L_0$ acts on the generating
function \eqref{gen_fn} (it should be considered as
function of $\kappa$; $z$ is an external parameter)  in the following way
\begin{equation}
L_0[\hat{f}^{(\kappa)}]=z\frac{d}{dz}\,\hat{f}^{(\kappa)}(z).
\label{idL}
\end{equation}
From this one can easily obtain a formal expression for the kernel
of the operator $L_0$
\begin{equation}
L_0(\kappa,\kappa')\sim\sum_{n=1}^{\infty}nv^{(\kappa)}_nv^{(\kappa')}_n.
\label{LK}
\end{equation}
It was argued in \cite{gm} that this kernel is not defined as
a distribution.
This means that the operator $L_0$ is not a kernel operator in the $\kappa$-basis.

It was explained in the Introduction that operator $L_0$ is completely
determined by its restriction onto the one-particle Fock space.
To actually find it we will use the following
strategy. First, find the inverse of the operator $L_0$ on one
particle Fock subspace. In other words we are looking for
an operator $G_0$ such that
\begin{equation*}
G_0L_0|\psi\rangle=L_0G_0|\psi\rangle=|\psi\rangle
\end{equation*}
for any state $|\psi\rangle=a^{\dag}(f)|0\rangle$ specifying
by function $f$ from Schwartz space. Using relation \eqref{LaLf}
one can formulate this statement in the Schwartz space as
\begin{equation}
G_0[L_0[f]](\kappa)=L_0[G_0[f]](\kappa)=f(\kappa)
\label{GLf}
\end{equation}
for any Schwartz function $f(\kappa)$.
From \eqref{LK} one can easily obtain a formal expression for
the kernel of operator $G_0$
\begin{equation}
G_0(\kappa,\kappa')=\sum_{n=1}^{\infty}\frac{1}{n}\,v^{(\kappa)}_nv^{(\kappa')}_n.
\end{equation}
Straight forward calculations of this kernel, which
I present in Appendix~\ref{app:A},
show that operator $G_0$ is indeed a kernel operator and
\begin{equation}
G_0(\kappa,\kappa')=\left[\frac{\theta(\kappa)}{\kappa}\right]^{1/2}
\left[\frac{\theta(\kappa')}{\kappa'}\right]^{1/2}\,
\frac{1}{4\cosh\left[\frac{\pi}{4}(\kappa-\kappa')\right]}.
\label{L0_1}
\end{equation}
Here $\theta(\kappa)=2\tanh\frac{\pi\kappa}{4}$ is a non-commutativity
parameter specifying the continuous Moyal algebra \cite{gm}.

Second, we find the inverse of operator $G_0$ on
the one-particle Fock space. In other words we are going to
solve equation \eqref{GLf}. The resulting operator will
coincide with the operator $L_0$ on the one-particle Fock space.
But $L_0$ is completely determined by its action
on this subspace and therefore we will actually find
operator $L_0$ on the whole Fock space.

So we need to solve the equation $G_0[L_0[f]]=f$:
\begin{equation}
\int_{-\infty}^{\infty}d\kappa'\,
\left[\frac{\theta(\kappa)}{\kappa}\right]^{1/2}
\,
\frac{1}{4\cosh\left[\frac{\pi}{4}(\kappa-\kappa')\right]}\,
\left[\frac{\theta(\kappa')}{\kappa'}\right]^{1/2}\,
g(\kappa')=f(\kappa),
\label{fLg}
\end{equation}
where $g=L_0[f]$.
It is convenient to introduce new functions $G$ and  $F$ via
\begin{equation}
G(\kappa')=\left[\frac{\theta(\kappa')}{\kappa'}\right]^{1/2}\,
g(\kappa')
\quad\text{and}\quad
F(\kappa)=\left[\frac{\theta(\kappa)}{\kappa}\right]^{-1/2}\,f(\kappa).
\label{fFgG}
\end{equation}
Then  equation \eqref{fLg} takes a from
\begin{equation}
\int_{-\infty}^{\infty}d\kappa'\,
\frac{1}{4\cosh\left[\frac{\pi}{4}(\kappa-\kappa')\right]}\,
G(\kappa')=F(\kappa).
\label{FFGG}
\end{equation}
This equation can be easily solved by using Fourier transformation method:
\begin{equation*}
F(\kappa')=\frac{1}{2\pi}\int_{-\infty}^{\infty}d\omega\,e^{i\omega\kappa'}\tilde{F}(\omega)
\end{equation*}
Using that
\begin{equation*}
\int_{-\infty}^{\infty}d\kappa'\,
\frac{e^{i\omega\kappa'}}{4\cosh\left[\frac{\pi}{4}(\kappa-\kappa')\right]}
=\frac{e^{i\omega\kappa}}{\cosh 2\omega}
\end{equation*}
one obtains
\begin{equation*}
\tilde{G}(\omega)=\cosh 2\omega\,\tilde{F}(\omega).
\end{equation*}
Finally the solution to \eqref{FFGG} is of the form
\begin{equation}
G(\kappa)=\frac{1}{2\pi}\int_{-\infty}^{\infty}
d\omega\int_{-\infty}^{\infty}
d\kappa'\,e^{i\omega(\kappa-\kappa')}\cosh 2\omega\,F(\kappa').
\label{f1}
\end{equation}
The integral over $\omega$ does not exist (this
once again confirms that $L_0$ is not a kernel
operator) and therefore  we can not simply
switch integrals over $\omega$ and over $\kappa'$. One can easily
see that the integral operator in the right hand side is
well defined at least on functions on $\kappa'$ with Gaussian
fall off at infinity.
One can give a precise mathematical meaning to this operator
if instead of Fourier transform one will use a Laplace transform
and define $0<\kappa<\infty$. In this
case we can rigorously prove that\footnote{Of course,
expression \eqref{FG} can be obtained by a formal
integration of \eqref{f1}.}
\begin{equation}
G(\kappa)=\frac{1}{2}\Bigl[F(\kappa+2i-i0)+F(\kappa-2i+i0)\Bigr].
\label{FG}
\end{equation}
The function $F$ here is supposed to be holomorphic
function in the strip $-2<\Im \kappa<2$. It
is allowed to have poles on the boundary of the strip.
The $\pm i0$ in the formula above tells us how one should approach this poles.
Namely one has to approach the pole on the line $\kappa+2i$ from the bottom and
the one on the
line $\kappa-2i$ from the top.
Substitution of \eqref{fFgG} into \eqref{FG} yields expression
\eqref{mainL0}
for the operator $L_0$.

Let us now check that \eqref{FG}
is indeed a solution to equation \eqref{FFGG}:
\begin{multline*}
\int_{-\infty}^{\infty}d\kappa'\,
\frac{1}{4\cosh\left[\frac{\pi}{4}(\kappa-\kappa')\right]}\,
\frac{1}{2}\Bigl[F(\kappa+2i-i0)+F(\kappa-2i+i0)\Bigr]
\\
=
\frac{1}{8}\int_{-\infty}^{\infty}d\kappa' F(\kappa')
\left[
\frac{i}{\sinh\frac{\pi}{4}(\kappa-\kappa')+i\varepsilon\cosh\frac{\pi}{4}(\kappa-\kappa')}
+\frac{-i}{\sinh\frac{\pi}{4}(\kappa-\kappa')-i\varepsilon\cosh\frac{\pi}{4}(\kappa-\kappa')}
\right]
\\
=F(\kappa).
\end{multline*}
Here in the middle line we have made a shift of integration variable, which
is allowed due to the fact that $F(\kappa)$ is holomorphic function on the strip
$-2<\Im\kappa<2$. This completes the derivation of formula \eqref{mainL0}
defining the operator $L_0$ in the class of difference operators.

Let me show now that the operator $L_0$ defining through \eqref{mainL0}
acts onto generating function $f^{(\kappa)}$ in the way
it requires by \eqref{idL}. Indeed,
\begin{multline}
L_0[f^{(\kappa)}(z)]=\frac{1}{4}\left[
\frac{\sqrt{\kappa(\kappa+2i)}}{(\kappa+2i)\sqrt{\frac{2}{\kappa+2i}
\sinh(\frac{\pi\kappa}{2})e^{i\pi}}}\,
\Bigl(1-e^{-(\kappa+2i)\arctan z}\Bigr)
\right.
\\
\left.
+\frac{\sqrt{\kappa(\kappa-2i)}}{(\kappa-2i)\sqrt{\frac{2}{\kappa-2i}
\sinh(\frac{\pi\kappa}{2})e^{-i\pi}}}\,
\Bigl(1-e^{-(\kappa-2i)\arctan z}\Bigr)
\right]
\\
=\frac{i}{4}\,e^{-\kappa\arctan z}\,\bigl(e^{-2i\arctan z}-e^{2i\arctan z}\bigr)
=\frac{z}{1+z^2}\,e^{-\kappa\arctan z}
=z\frac{d}{dz}\,f^{(\kappa)}(z).
\label{L0fk}
\end{multline}


\section{Operators $L_1$ and $L_{-1}$}
\label{sec:L1}
\setcounter{equation}{0}

\dopage{\finkfile}

The operators $L_1$ and $L_{-1}$ are related by hermitian
conjugation. Therefore it is enough to find the operator $L_{-1}$.
As it is done in the previous section
we restrict our considerations
to the zero momentum sector. The action of $L_{-1}$ on the generating
function $f^{(\kappa)}(z)$ is given by the formula
\begin{equation}
L_{-1}[f^{(\kappa)}(z)]=-\frac{1}{\Nc(\kappa)}+\frac{d}{dz}\,f^{(\kappa)}(z).
\end{equation}
Therefore formally its kernel can be written as
\begin{equation}
L_{-1}(\kappa,\kappa')\sim\sum_{m=1}^{\infty}v_{m+1}^{(\kappa)}
\,\sqrt{(m+1)m}\,v_{m}^{(\kappa')}
\label{kerL_1}
\end{equation}
As it is explained in the Introduction the operator $L_{-1}$ is
completely determined by its action onto the one-particle Fock space.
Therefore in order to find it we will use the same method as the one
used in the previous section. There will be only one
modification of the method related to the fact
that the operator $L_1$ (conjugated to the operator $L_{-1}$)
has a one-dimensional kernel. First, we find an operator
$G_{-1}$ such that it ``inverts'' $L_{-1}$ on the one-particle Fock space.
More precisely $G_{-1}$ satisfies the equation
\begin{subequations}
\begin{equation}
G_{-1}[L_{-1}[f]](\kappa)=f(\kappa)\quad
\text{and}\quad
L_{-1}[G_{-1}[f]](\kappa)=P_1[f](\kappa),
\label{L_1G}
\end{equation}
where $P_1$ is a projector on the subspace on which $L_1$ is
a non-degenerate operator:
\begin{equation}
P_{1}[f](\kappa)=f(\kappa)-v_{1}^{(\kappa)}\,\int_{-\infty}^{\infty}
d\kappa'\,v_{1}^{(\kappa')}f(\kappa').
\end{equation}
\end{subequations}
From equation \eqref{kerL_1} one can easily obtain the formal
expression for the kernel of operator $G_{-1}$:
\begin{equation}
G_{-1}(\kappa,\kappa')=
\sum_{m=1}^{\infty}v_{m}^{(\kappa)}
\frac{1}{\sqrt{m(m+1)}}\,v_{m+1}^{(\kappa')}.
\end{equation}
Straightforward calculations (very similar to the ones
presented in Appendix~\ref{app:A}) show that this expression
does indeed define a distribution:
\begin{equation}
G_{-1}(\kappa,\kappa')=
\frac{1}{\kappa\sqrt{\Nc(\kappa)\Nc(\kappa')}}
-\frac{1}{2\kappa}\sqrt{\frac{\Nc(\kappa)}{\Nc(\kappa')}}\,
\frac{\kappa-\kappa'}{\sinh\frac{\pi}{2}(\kappa-\kappa')}
\end{equation}
and therefore $G_{-1}$ is a kernel operator.

Second, we find the ``inverse'' of operator $G_{-1}$ on
the one particle Fock space. In other words we are going to
solve the first equation in \eqref{L_1G}. The resulting operator will
coincide with the operator $L_{-1}$ on the one-particle Fock space.
Since $L_{-1}$ is completely determined by its action
on this subspace  we will actually find
operator $L_{-1}$ on the whole Fock space.

So we need to solve the equation
\begin{equation}
\int_{-\infty}^{\infty}d\kappa'\,
\frac{1}{\kappa\sqrt{\Nc(\kappa)\Nc(\kappa')}}\,g(\kappa')
-\frac{1}{2\kappa}\int_{-\infty}^{\infty}d\kappa'\,
\sqrt{\frac{\Nc(\kappa)}{\Nc(\kappa')}}\,
\frac{\kappa-\kappa'}{\sinh\frac{\pi}{2}(\kappa-\kappa')}
\,g(\kappa')=f(\kappa),
\end{equation}
where $g(\kappa')=L_{-1}[f](\kappa')$. Notice now
that by the construction $g$ is a function that belongs
to the image of $L_{-1}$, therefore it satisfies the equation $P_1[g]=g$.
This equation just reflects the fact that there is no state $a_1^{\dag}|0\rangle$
in the image of operator $L_{-1}$ restricted to the one-particle subspace.
Eventually the first term in the equation is identically zero\footnote{
We use the fact that $v^{(\kappa')}_1=\frac{1}{\sqrt{\Nc(\kappa')}}$}.
It is useful to introduce new functions $F$ and $G$
\begin{equation}
G(\kappa')=\sqrt{\Nc(\kappa')}\,g(\kappa')
\quad\text{and}\quad
F(\kappa)=\kappa\sqrt{\Nc(\kappa)}\,f(\kappa)
\label{L1FG}
\end{equation}
for which the equation takes extremely simple form
\begin{equation*}
-\frac{1}{2}\int_{-\infty}^{\infty}d\kappa'\,
\frac{\kappa'-\kappa}{\sinh\frac{\pi}{2}(\kappa'-\kappa)}\,G(\kappa')
=F(\kappa).
\end{equation*}
This equation can be solved using method of Fourier transform.
Using that
\begin{equation*}
\int_{-\infty}^{\infty}d\kappa'\,e^{i\omega\kappa'}
\frac{\kappa'}{\sinh\frac{\pi}{2}\kappa'}
=\frac{2}{\cosh^2\omega}
\end{equation*}
one can easily obtain the solution
\begin{equation*}
G(\kappa)=-\frac{1}{2}\,F(\kappa)-\frac{1}{4\pi}
\int_{-\infty}^{\infty}d\omega\int_{-\infty}^{\infty}d\kappa'\,
e^{i\omega(\kappa-\kappa')}\cosh 2\omega F(\kappa').
\end{equation*}
Using the arguments we presented equation \eqref{f1} one obtains
\begin{equation*}
G(\kappa)=-\frac{1}{2}\,F(\kappa)-\frac{1}{4}\Bigl[
F(\kappa+i2^-)+F(\kappa-i2^-)
\Bigr].
\end{equation*}
Here the function $F(\kappa)$ is supposed to be holomorphic
on the strip $-2<\Im\kappa<2$. Now using the relations
\eqref{L1FG} one obtains expression \eqref{mainL1}
defining the operator $L_{-1}$ in a class
of difference operators. An expression for the operator
$L_1$ is easily obtained by hermitian conjugation.

Using the same technique as we used in deriving \eqref{L0fk} we can
show that operators $L_1$ and $L_{-1}$ defined via \eqref{mainL1}
act on the generating function \eqref{gen_fn} as follows
\begin{subequations}
\begin{align}
L_1[f^{(\kappa)}(z)]&=z^2\frac{d}{dz}\,f^{(\kappa)}(z)
\\
L_{-1}[f^{(\kappa)}(z)]&=-\frac{1}{\Nc(\kappa)}+\frac{d}{dz}\,f^{(\kappa)}(z)
\end{align}
\label{L1L_1}
\end{subequations}


\section{Commutation relations}
\label{sec:com}
\setcounter{equation}{0}

\dopage{\finkfile}
In this section we are going to show that
operators $L_0$, $L_{\pm 1}$ represented by
the difference operators \eqref{mainL0} and \eqref{mainL1}
satisfy the standard commutation relations.

First, let me demonstrate how to calculate the commutator of $L_1$ and $L_{-1}$.
From \eqref{mainL1} we obtain the identities
\begin{subequations}
\begin{align}
L_1[L_{-1}[g]](\kappa)&=-\frac{\kappa}{2}\,L_{-1}[g]+\frac{i}{4}\left[
\sqrt{\kappa(\kappa+2i)}L_{-1}[g](\kappa+2i)
-\sqrt{\kappa(\kappa-2i)}\,L_{-1}[g](\kappa-2i)
\right];
\\
L_{-1}[L_1[g]](\kappa)&=-\frac{\kappa}{2}\,L_{1}[g]-\frac{i}{4}\left[
\sqrt{\kappa(\kappa+2i)}L_{1}[g](\kappa+2i)
-\sqrt{\kappa(\kappa-2i)}\,L_{1}[g](\kappa-2i)
\right].
\end{align}
\end{subequations}
By subtracting the last equation from the first one one obtains
\begin{multline*}
[L_1,\,L_{-1}][g](\kappa)=\frac{\kappa}{2}\bigl(L_1[g]-L_{-1}[g]\bigr)(\kappa)
\\
+\frac{i}{4}\left[
\sqrt{\kappa(\kappa+2i)}\bigl(L_1[g]+L_{-1}[g]\bigr)(\kappa+2i)
-\sqrt{\kappa(\kappa-2i)}\bigl(L_1[g]+L_{-1}[g]\bigr)(\kappa-2i)
\right]
\end{multline*}
Substitution of \eqref{mainL1} and simplification yield
\begin{equation}
[L_1,L_{-1}][g]=2L_0[g].
\end{equation}

Second , we check commutation relations between $L_0$
and $L_1$.
\begin{multline}
[L_0,\,L_1][g](\kappa)=\frac{\kappa}{2}L_0[g](\kappa)
\\
+\frac{1}{4}\left[
\sqrt{\kappa(\kappa+2i)}\bigl(L_1[g]-iL_0[g]\bigr)(\kappa+2i)
+\sqrt{\kappa(\kappa-2i)}\bigl(L_1[g]+iL_0[g]\bigr)(\kappa-2i)
\right]
\end{multline}
Substitution of \eqref{mainL0} and \eqref{mainL1} and simplification yield
\begin{equation}
[L_0,\,L_1][g]=-L_1[g].
\end{equation}


\section*{Acknowledgments}
I would like to thank A.~Konechny, H.~Liu, S.~Lukyanov
and G.~Moore for useful discussions.
I would like to thank H.~Liu and G.~Moore for showing me
the draft of paper \cite{gm}. I would like to thank
A.~Konechny and M.~Kroyter for the comments on this paper.
The work of  was supported in part by RFBR grant 02-01-00695.

\vspace{1.5cm}
\textbf{Note added:}
While this paper was nearing completion, the paper \cite{0210155} appeared,
which contains in Section~5 similar expressions for $L_0$ and $L_{\pm 1}$.
The reason for publishing this paper is mainly to give another
point of view on the Virasoro operators in the $\kappa$-basis.

\clearpage
\appendix
\section*{Appendix}
\addcontentsline{toc}{section}{Appendix}
\renewcommand {\theequation}{\thesection.\arabic{equation}}

\section{Calculations of $G_0$}
\label{app:A}
\setcounter{equation}{0}

\dopage{\finkfile}

The kernel for operator $G_0$ can be written as the
following contour integral
\begin{equation}
G_0(\kappa,\kappa')=\frac{1}{\kappa\kappa'\sqrt{\Nc(\kappa)\Nc(\kappa')}}
\frac{1}{2\pi i}\,
\oint_C \frac{dz}{z}\,(1-e^{-\kappa\arctan\frac{1}{z}})(1-e^{-\kappa'\arctan z}).
\end{equation}
Notice that this integral has two logarithmic singularities at the points
$\pm i$. Therefore the complex plane has two cuts starting at those points.
We choose them to lie on the imaginary axis (for more details
on the contour see Appendix in \cite{0207001}). To calculate
the integral we deform the contour $C$ in such way that it will
 go along  the cuts.
We introduce $z=ix-\epsilon$,
 $z=ix+\epsilon$, $z=-ix+\epsilon$ and $z=-ix-\epsilon$
 on contours $C_+$, $C_-$, $C_+'$ and $C_-'$ respectively
\begin{align*}
\kappa\kappa'\sqrt{\Nc(\kappa)\Nc(\kappa')}\,G_0(\kappa,\kappa')=
\frac{1}{2\pi i}\int_1^{\infty}
&\left[\frac{idx}{ix-\epsilon}\,
\Bigl(1-e^{-\kappa\arctan\frac{1}{ix-\epsilon}}\Bigr)
\Bigl(1-e^{-\kappa'\arctan(ix-\epsilon)}\Bigr)
\right.
\\
&+\frac{-idx}{ix+\epsilon}\,
\Bigl(1-e^{-\kappa\arctan\frac{1}{ix+\epsilon}}\Bigr)
\Bigl(1-e^{-\kappa'\arctan(ix+\epsilon)}\Bigr)
\\
&+\frac{-idx}{-ix+\epsilon}\,
\Bigl(1-e^{-\kappa\arctan\frac{1}{-ix+\epsilon}}\Bigr)
\Bigl(1-e^{-\kappa'\arctan(-ix+\epsilon)}\Bigr)
\\
&\left.
+\frac{idx}{-ix-\epsilon}\,
\Bigl(1-e^{-\kappa\arctan\frac{1}{-ix-\epsilon}}\Bigr)
\Bigl(1-e^{-\kappa'\arctan(-ix-\epsilon)}\Bigr)
\right]
\end{align*}
Using that $\arctanh(ix\pm\epsilon)=\pm\frac{\pi}{2}+i\coth^{-1} x$ and
$\arctanh(ix\pm\epsilon)^{-1}=-i\coth^{-1} x$
one obtains
\begin{align*}
=\frac{1}{2\pi i}\int_1^{\infty}\frac{dx}{x}
&\left[
\Bigl(1-e^{-\kappa(-i\coth^{-1}x)}\Bigr)
\Bigl(1-e^{-\kappa'(-\frac{\pi}{2}+i\coth^{-1}x)}\Bigr)
\right.
\\
&-
\Bigl(1-e^{-\kappa(-i\coth^{-1}x)}\Bigr)
\Bigl(1-e^{-\kappa'(\frac{\pi}{2}+i\coth^{-1}x)}\Bigr)
\\
&+
\Bigl(1-e^{-\kappa i\coth^{-1}x}\Bigr)
\Bigl(1-e^{-\kappa'(\frac{\pi}{2}-i\coth^{-1}x)}\Bigr)
\\
&\left.
-
\Bigl(1-e^{-\kappa i\coth^{-1}x}\Bigr)
\Bigl(1-e^{-\kappa'(-\frac{\pi}{2}-i\coth^{-1}x)}\Bigr)
\right]
\end{align*}
Now the change of variable
\begin{equation*}
x=\coth u\quad\text{and}\quad dx=-\frac{du}{\sinh^2 u}
\end{equation*}
yields
\begin{align*}
=\frac{1}{\pi i}\int_0^{\infty}\frac{du}{\sinh 2u}
&\left[
\Bigl(1-e^{i\kappa u}\Bigr)
\Bigl(1-e^{\frac{\pi\kappa'}{2}-i\kappa' u}\Bigr)
\right.
\\
&-
\Bigl(1-e^{i\kappa u}\Bigr)
\Bigl(1-e^{-\frac{\pi\kappa'}{2}-i\kappa' u}\Bigr)
\\
&+
\Bigl(1-e^{-i\kappa u}\Bigr)
\Bigl(1-e^{-\frac{\pi\kappa'}{2}+i\kappa' u}\Bigr)
\\
&\left.
-
\Bigl(1-e^{-i\kappa u}\Bigr)
\Bigl(1-e^{\frac{\pi\kappa'}{2}+i\kappa' u}\Bigr)
\right]
\end{align*}
Simplification yields
\begin{multline}
=\frac{2}{\pi i}\,\sinh\frac{\pi\kappa'}{2}
\int_0^{\infty}\frac{du}{\sinh 2u}\,\left[
e^{i\kappa' u}-e^{i(\kappa'-\kappa)u}
-e^{-i\kappa' u}+e^{-i(\kappa'-\kappa)u}
\right]
\\
=
\frac{2}{\pi i}\,\sinh\frac{\pi\kappa'}{2}
\int_{-\infty}^{\infty}du\,\mathscr{P}\frac{1}{\sinh 2u}\,\left[
e^{i\kappa' u}+e^{-i(\kappa'-\kappa)u}
\right]
\end{multline}
Using the fact that
\begin{equation}
\frac{1}{\pi i}
\int_{-\infty}^{\infty}du\,\mathscr{P}\frac{1}{\sinh 2u}\,e^{i\beta u}
=\frac{1}{2}\,\tanh\frac{\pi\beta}{4}
\end{equation}
one obtains
\begin{equation}
=\sinh\frac{\pi\kappa'}{2}\left[
\tanh\frac{\pi\kappa'}{4}+\tanh\frac{\pi(\kappa-\kappa')}{4}
\right]=2\,\frac{\sinh\frac{\pi\kappa}{4}\sinh\frac{\pi\kappa'}{4}}
{\cosh\frac{\pi(\kappa-\kappa')}{4}}.
\end{equation}
Finally one gets
\begin{equation}
G_0(\kappa,\kappa')=\left[\frac{\theta(\kappa)}{\kappa}\right]^{1/2}
\left[\frac{\theta(\kappa')}{\kappa'}\right]^{1/2}\,
\frac{1}{4\cosh\left[\frac{\pi}{4}(\kappa-\kappa')\right]}.
\end{equation}


\clearpage

\end{document}